\documentclass[final]{IEEEtran}

\usepackage{amsthm,amssymb,graphicx,multirow,amsmath,color,amsfonts}%,ulem}
\usepackage[update,prepend]{epstopdf}
\usepackage[noadjust]{cite}
\usepackage[latin1]{inputenc}
\usepackage{tikz}
\usetikzlibrary{arrows,calc}		% Optional ticks libraries
\usepackage{bbm} % for \mathbbm{1}
\usepackage{pdfpages}
\usepackage{tabulary}
\usepackage{multirow}
\usepackage{comment}
\usepackage{cite}
\def\sir{\mathtt{SIR}}

\theoremstyle{definition}

\usepackage{hyperref}

\def\nb0{{\mathbf{0}}}
\def\nb1{{\mathbf{1}}}

\newtheorem{lemma}{Lemma}

\newtheorem{ndef}{Definition}

\newtheorem{theorem}{Theorem}

\newtheorem{cor}{Corollary}

\def\sir{\mathtt{SIR}}

\allowdisplaybreaks % Allows breaking of eqnarray over multiple pages (avoids unnecessary blanks in the document before eqnarray)

\begin{document}
\graphicspath{{./Figures/}}
\title{Modeling and Performance Analysis of Full-Duplex Communications in Cache-Enabled D2D Networks}
\author{Mansour Naslcheraghi, Mehrnaz Afshang, Harpreet S. Dhillon
	
	\thanks{M. Naslcheraghi is with Cognitive Radio Laboratory, Department of ECE, Shahid Beheshti University, Tehran, Iran (email: m.naslcheraghi@ieee.org)} 
	
	\thanks{M. Afshang and H. S. Dhillon  are with Wireless@VT, Department of ECE, Virginia Tech, Blacksburg, VA (email: \{mehrnaz, hdhillon\}@vt.edu).} 
}

\maketitle
\thispagestyle{empty}
\pagestyle{empty}

\begin{abstract}
	Cache-enabled Device-to-Device (D2D) communication is widely recognized as one of the key components of the emerging fifth generation (5G) cellular network architecture. However, conventional half-duplex (HD) transmission may not be sufficient to provide fast enough content delivery over D2D links in order to meet strict latency targets of emerging D2D applications. In-band full-duplex (FD), with its capability of allowing simultaneous transmission and reception, can improve spectral efficiency and reduce latency by providing more content delivery opportunities. In this paper, we consider a finite network of D2D nodes in which each node is endowed with FD capability. We first carefully list all possible operating modes for an arbitrary device using which we compute the number of devices that are actively transmitting at any given time. We then characterize network performance in terms of the success probability, which depends on the content availability, signal-to-interference ratio ($\sir$) distribution, as well as the operating mode of the D2D receiver. Our analysis concretely demonstrates that caching dictates the system performance in lower target $\sir$ thresholds whereas interference dictates the performance at the higher target $\sir$ thresholds. 
\end{abstract}


\section{Introduction} \label{sec:intro}
In the 5G evolution, cache-enabled D2D technology has attracted widespread attention because of its potential to improve system performance and enhance user experience \cite{Negin_Mag}. The main idea behind this technology is to use the local storage of the user devices to store popular content and deliver it asynchronously to proximate devices through D2D communications whenever they need it. Some early results on the fundamental limits of wireless D2D caching systems appear in \cite{Fundamental_caching} in which the authors developed constructive achievability coding strategies and information theoretic bounds for a D2D caching network under the constraint of arbitrary demands. Further, one of the schemes being proposed for the 5G mobile communications systems is FD communication, which allows simultaneous transmission and reception on the same channel. In a cache-enabled D2D networks, FD radios can promise more advantages in comparison with its HD counterpart \cite{MyIET,myPPP}, by providing more content delivery opportunities, thus improving spectral efficiency and reducing end-to-end delay. 

The accurate performance analysis of the cache-enabled D2D communications requires consideration of the caching mechanism not only in the caching performance, but also in the $\sir$ distribution. More precisely, the caching mechanism determines the possible transmitting users and the user operating modes (HD/FD) for the receiver of interest which directly impacts $\sir$. This consideration leads to more challenging analysis compared to the case in which the $\sir$ distribution is assumed to be independent of the caching mechanism due to two main reasons. First, the performance depends upon the D2D network formation, which in turn depends upon the content cached in the user devices as well as their demands. For instance, it is reasonable to say that two users will initiate D2D link only if at least one of them finds its desired content in the other's cache, and the experienced $\sir$ at the receiver exceeds some predefined target $\sir$ threshold. Second, depending upon the cached content and the user demand, an arbitrary node can operate in either HD or FD mode. Even when an arbitrary node operates in FD, it does not necessarily form the more intuitive bi-directional FD (BFD) link in which two devices exchange data with each other. Another possibility is a three node FD (TNFD) collaboration in which an intermediate node can receive its desired content from one node and concurrently serve some other node using content stored in its cache. These challenges that appear in the analysis of cache-enabled D2D networks with FD capability have not yet been addressed and will form the basis of our contribution.

As evident from the above discussion already, an arbitrary user can operate in different {\em operating modes} depending upon the cached content and random demands. All possible operating modes are illustrated in Fig. \ref{Fig: D2DCasesGraph}. and will be discussed in more detail in the next Section. In the existing literature, e.g., see~\cite{cacherequestPPP, HitProbe1, HitProbe2, HitProbe3,malak,MehrnazBPP}, the focus has mostly been on the performance analysis of an arbitrary node when it is obtaining content from a proximate node. This implicitly means that the receiver of interest is always assumed in half-duplex mode which corresponds to the half-duplex receiver (HDRX) case in Fig. \ref{Fig: D2DCasesGraph}. As will be discussed in the sequel, the random operating modes also affect the $\sir$ distribution. Therefore, the existing works focus on a very specific case out of all possible scenarios that could occur in a D2D network in which the nodes are endowed with FD capability. In this paper, we overcome this shortcoming by modeling these operating modes and their impact on the system performance accurately. More details about the main contributions are provided next. 

\textit{Contributions}. 
In this paper, we consider a cache-enabled D2D network formed by a fixed number of users whose locations are modeled as a Binomial Point Process (BPP). We carefully list all possible operating modes when users have the FD capability. We then derive closed form expressions for the probabilities that an arbitrarily selected user is operating in one of these modes. These probabilities are used to derive the probability mass function (PMF) of the number of nodes that actively transmit at any given time. Using this PMF, we then characterize the $\sir$ distribution for the HD/FD receiver of interest in the presence of power control as well as the success probability for an arbitrary node.

%%%%%%%%%%%%%%%%%%%%%%
\section{System Model} \label{sec:SysMod}
%%%%%%%%%%%%%%%%%%%%%%
We consider a finite network consisting of fixed $N$ number of users forming a BPP inside a disk ${\rm b}(\textbf{o},\mathcal{R})\subset \mathbb{R}^2$ with radius $\mathcal{R}$. Users are assumed to have the capability of FD communication. In other words, these nodes are assumed to be located uniformly at random independently of each other over the disk. Denoting by $\{{\textbf{y}}_i \} \equiv \Phi$ the locations of the users, the probability density function (PDF) of each element ${\textbf{y}}_i$ is
\begin{align}
f(\textbf{y}_i)=
\begin{cases}
\frac{1}{\pi \mathcal{R}^2} &;\left\| \textbf{y}_i\right\| \le \mathcal{R},\\
0 &;\textup{o.w}.
\end{cases}.
\end{align}

\subsection{Caching Model}
\label{subsec:CachingModel}
%%%%%%%%%%%%%%%%%%%%%%%%%%%%
Denote the library of popular contents of size $m$ by $\mathbf{L} = \{{c_\ell}\}_{\ell=1:m}$. Each content has an associated popularity score, which is characterized by the user requests. Each user has a unique identity $u_\kappa$, $\kappa \in \{1,2,\dots N\}$. To determine which contents are cached in each user device, we use optimal caching policy \cite{Negin_Mag,MPC-Paper}. According to this policy, contents are pushed in the user devices in advance by the central base station (BS). Each content is associated to a single user, which means that there is no overlap between cached contents in user devices. While each user has the capability of storing multiple contents, for the sake of simplicity we assume that each user caches one content. Under these assumptions, user $u_\kappa$ is assumed to cache content $c_\kappa$, where $c_{\kappa}$ is different across users. The case of caching multiple contents is left as a promising direction for future work. The popularity of content $c_\kappa$ is equivalent to the probability of requesting content $c_\kappa$. The request probability is  denoted by $\rho_\kappa$ and defined by
\begin{equation}
\label{Formula:PopularityDist}
{\rho_\kappa} = \Upsilon(\kappa,\gamma_r,m),
\end{equation}
where $\Upsilon(.)$ is the popularity distribution, and the parameter $\gamma_r$ is the skew exponent and characterizes the popularity distribution by controlling popularity of the contents for a given library size $m$. Each user randomly requests a content from the library according to popularity distribution given by (\ref{Formula:PopularityDist}). A pair of users can potentially initiate a D2D connection if one of them finds its desired content in the other user. Based on the information of the cached contents and users' requests, there will be different operating modes for an arbitrary user, which are introduced next.
\begin{figure}[t]
	\centering
	\includegraphics[width=0.26 \textwidth]{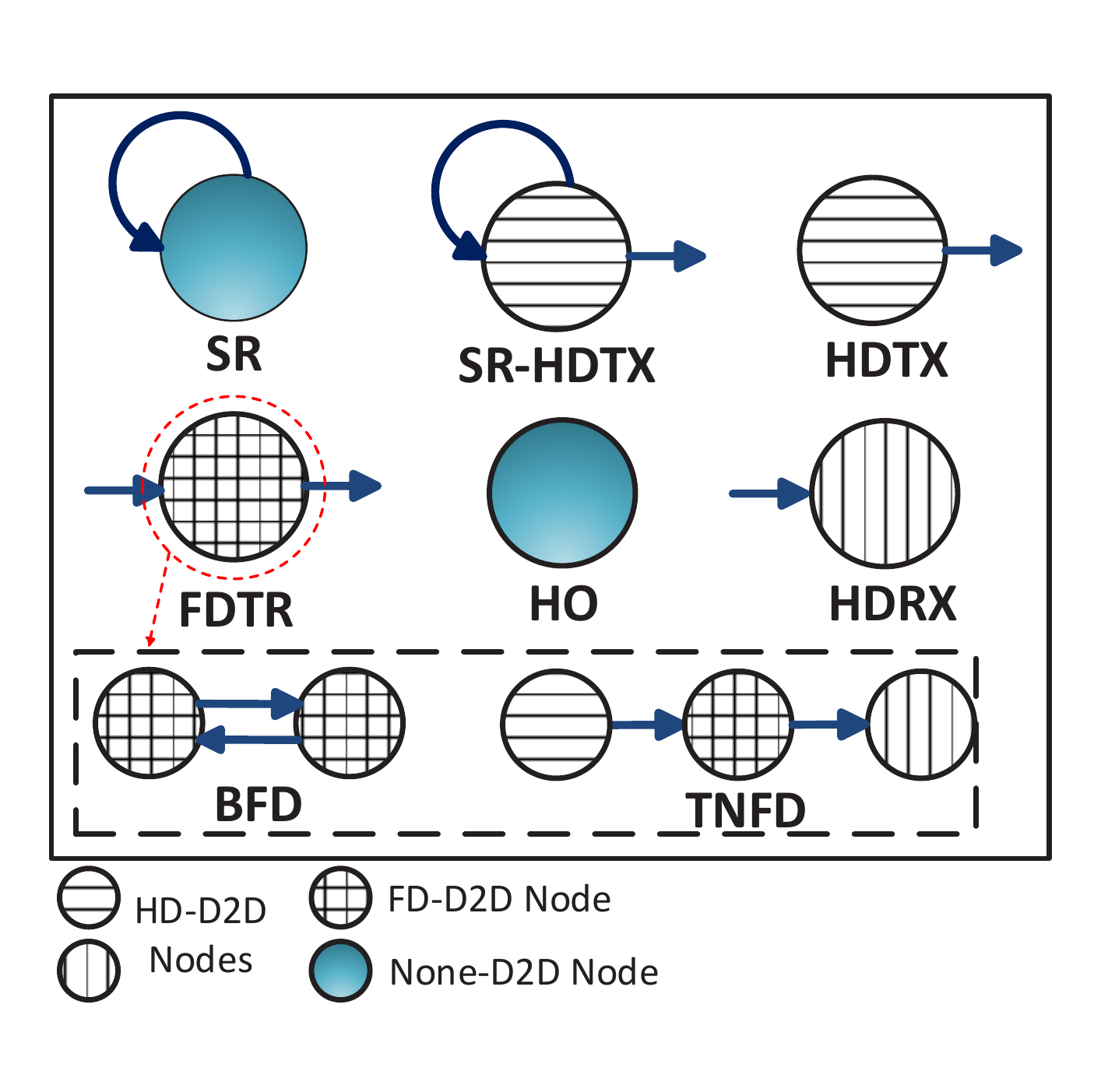}%\vspace{-6mm}
	\caption{All possible operating modes for an arbitrary node.} 
	\label{Fig: D2DCasesGraph}
\end{figure}
\begin{figure}[t]
	\centering
	\includegraphics[width=0.24 \textwidth]{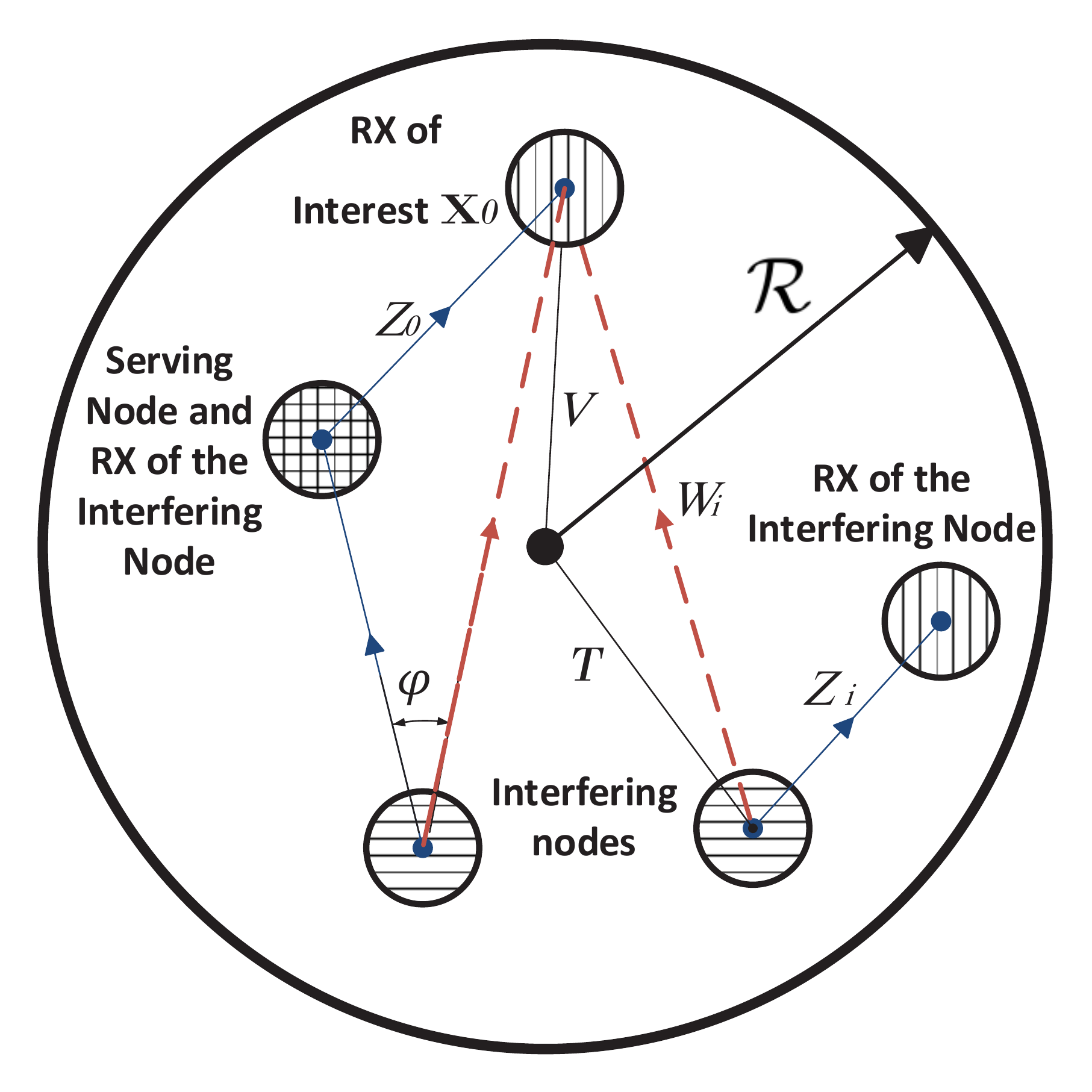}%\vspace{-4mm}
	\caption{Illustration of the involved distances in the analysis.}
	\label{Fig:RelevantDistances}
\end{figure}

%%%%%%%%%%%%%%%%%%%%%%%%%%%%%%%%%%
\subsection{Modeling User Operating Modes}
\label{subsec:UserOperatingModes}
%%%%%%%%%%%%%%%%%%%%%%%%%%%%%%%%%%%
There are six different possible operating modes for an arbitrary node as shown in Fig. \ref{Fig: D2DCasesGraph}. Definitions of the operating modes are as follows. 
\begin{itemize}
	\item \textbf{Self-Request (SR)}: an arbitrary user can find its desired content in its own cache. We let $\mathcal{P}_{\textup{SR}}$ be the probability of this mode. 
	\item \textbf{Self-Request and HD Transmission (SR-HDTX)}: an arbitrary user can find its desired content in its own cache, and can concurrently serve for other users' demand. We let $\mathcal{P}_{\textup{SR-HDTX}}$ be the probability of this mode.
	\item \textbf{Full-Duplex Transceiver (FDTR)}: an arbitrary user can find its desired content in its vicinity through D2D link, and can concurrently serve for other users' demand. We let $\mathcal{P}_{\textup{FDTR}}$ be the probability of this mode. This case can be divided into two different configurations as follows: 
	\begin{itemize}
		\item \textbf{Bi-Directional Full-Duplex (BFD):} an arbitrary user can concurrently exchange content with another user. We let $\mathcal{P}_{\textup{BFD}}$ be the probability of this mode.
		\item \textbf{Three-Node Full-Duplex (TNFD):} an arbitrary user can concurrently receive and transmit from and to different user. We let $\mathcal{P}_{\textup{TNFD}}$ be the probability of this mode.
	\end{itemize} 
	\item \textbf{Half-Duplex Transmitter (HDTX)}: an arbitrary user cannot find its desired content either in its vicinity or in its own cache, however, it can serve for other users' demand. We let $\mathcal{P}_{\textup{HDTX}}$ be the probability of this mode.
	\item \textbf{Half-Duplex Receiver (HDRX)}: an arbitrary user can receive its desired content via D2D link, and there is no user(s) that demand(s) for the content that is cached in this user. We let $\mathcal{P}_{\textup{HDRX}}$ be the probability of this mode.
	
	\item \textbf{Hitting Outage (HO)}: an arbitrary user cannot find its desired content in its vicinity or its own cache, and there is no user(s) that demand(s) for the content that is cached in this user. We let $\mathcal{P}_{\textup{HO}}$ be the probability of this mode. 
\end{itemize}
One can say: $\mathcal{P}_{\rm FDTR} = \mathcal{P}_{\rm BFD} + \mathcal{P}_{\rm TNFD}$ and \\
$\mathcal{P}_{\rm SR} + \mathcal{P}_{\rm SR-HDTX} + \mathcal{P}_{\rm FDTR} + \mathcal{P}_{\rm HDTX} + \mathcal{P}_{\rm HDRX}+ \mathcal{P}_{\rm HO}=1$.

%%%%%%%%%%%%%%%%%%%%%%%%%%%
\subsection{Channel Model and System Key Assumptions}
\label{subsec:Channel Model}
%%%%%%%%%%%%%%%%%%%%%%%%%%%
In our system, all D2D pairs share the same time/frequency resources. Please refer to Fig. \ref{Fig:RelevantDistances} for an illustration of the network formation. For the channel model, standard power-law path loss model is considered in which the signal power decays at the rate of $r^{-\alpha}$, where $\alpha > 2$ is the path loss exponent. Independent Rayleigh fading with unit mean, i.e., $h_i \sim \exp(1)$ is assumed between any D2D pair. We employ full channel inversion power control, which means that the transmitting user completely compensates for the pathloss by increasing its transmit power by a factor of $Z_0^{\alpha}$, where $Z_0$ is the serving distance as shown in Fig. \ref{Fig:RelevantDistances}. We assume imperfect self-interference (SI) cancellation with residual power ratio $0 \le \beta \le 1$ for the FD radios in concurrent transmission and reception over the same time/frequency. We also assume that the background noise is negligible compared to the interference and is hence ignored. Let $\textbf{x}_0 \in {\rm b}(\textbf{o},\mathcal{R})$ be the location of an arbitrary receiver within the disk. Now, according to the D2D operating modes described in \ref{subsec:UserOperatingModes}, we have two types of receivers in the system: HDRX and FDTR. Letting $\textbf{y}_0$ be the location of the serving node, $\textbf{x}_0$ be the location of receiver of interest, $\left\|\textbf{y}_0 -\textbf{x}_0\right\|$ be the distance between the serving node and the receiver of interest, $\textbf{x}_i$ be the receiver of the interfering node, $\left\|\textbf{y}_i-\textbf{x}_0\right\|$ be the distance between the interfering node and the receiver of interest, and $\left\|\textbf{y}_i-\textbf{x}_i\right\|$ be the distance between the interfering node and its respective receiver located at $\textbf{x}_i \in {\rm b}({\bf o}, {\cal R})$, the $\sir$ at the receiver type $\delta \in \{\textup{HDRX}, \textup{FDTR}\}$, denoted by $\sir_{\delta}$, can be defined as
\begin{equation}
\label{Formula: SINR expression}
\sir_{\delta} = \frac{h_0}{\sum\nolimits_{\textbf{y}_i \in \Phi\backslash \textbf{y}_0} \left({h_i \left\|\textbf{y}_i-\textbf{x}_0\right\|^{-\alpha}}\left\|\textbf{y}_i-\textbf{x}_i\right\|^{\alpha} + \vartheta \right)},
\end{equation}
where  $\vartheta = \mathbbm{1}_\delta\beta \left\|\textbf{y}_0-\textbf{x}_0\right\|^{\alpha}$ and $\mathbbm{1}_\delta$ is an indicator function defined by ${\mathbbm{1}_\delta } = \left\{ \begin{array}{l}
\begin{array}{*{20}{c}}
1&{;\delta  = {\textup{FDTR}}}
\end{array}\\
\begin{array}{*{20}{c}}
0&{;\delta  = {\textup{HDRX.}}}
\end{array}
\end{array}\right.$

%%%%%%%%%%%%%%%%%%%%%%%%%%%%%%%%%
\section{Analyzing Operating Mode Probabilities} \label{Sec: Collaboration probability section}
%%%%%%%%%%%%%%%%%%%%%%%%%%%%%%%%%
In the following Theorem, we provide the closed form expressions for the probabilities of all possible operating modes described in subsection \ref{subsec:UserOperatingModes}.
\begin{theorem}
	\label{Theorem:OptimalCaching}
	The probabilities of all possible operating modes for an arbitrary user are
	\begin{flalign}
	&\mathcal{P}_{\textup{SR}}= \frac{1}{N} \sum_{\kappa=1}^{N} \rho_\kappa \left(1-\rho_\kappa\right)^{N-1}, \label{Formula: Probe SR} \\
	&\mathcal{P}_{\textup{SR-HDTX}} = \frac{1}{N} \sum_{\kappa=1}^{N}\rho_\kappa \left(1-\left(1-\rho_\kappa \right)^{N-1}\right), \label{Formula: Probe SR-HDTX}\\
	&\mathcal{P}_{\textup{FDTR}} = \frac{1}{N} \sum_{\kappa=1}^{N}\left(\mathcal{P}_{\textup{hit}}-\rho_\kappa \right)\left(1-\left(1-\rho_\kappa \right)^{N-1}\right), \label{Formula: Probe FDTR}\\
	&\mathcal{P}_{\textup{BFD}} = \frac{1}{N} \sum_{\kappa=1}^{N}\left(\mathcal{P}_{\textup{hit}}-\rho_\kappa \right)\rho_\kappa, \label{Formula: Probe BFD}\\
	&\mathcal{P}_{\textup{TNFD}} = \frac{1}{N} \sum_{\kappa=1}^{N}\left(\mathcal{P}_{\textup{hit}}-\rho_\kappa \right)\left(1-\rho_\kappa-\left(1-\rho_\kappa\right)^{N-1}\right), \label{Formula: Probe TNFD}\\
	&\mathcal{P}_{\textup{HDRX}} = \frac{1}{N} \sum_{\kappa=1}^{N}\left(\mathcal{P}_{\textup{hit}}-\rho_\kappa \right)\left(1-\rho_\kappa\right)^{N-1},\label{Formula: Probe HDRX}\\
	&\mathcal{P}_{\textup{HDTX}} = \frac{1}{N} \sum_{\kappa=1}^{N}\left(1-\mathcal{P}_{\textup{hit}}\right)\left(1-\left(1-\rho_\kappa \right)^{N-1}\right), \label{Formula: Probe HDTX}\\
	&\mathcal{P}_{\textup{HO}} = \frac{1}{N} \sum_{\kappa=1}^{N}\left(1-\mathcal{P}_{\textup{hit}}\right)\left(1-\rho_\kappa \right)^{N-1}, \label{Formula: Probe HO}
	\end{flalign}
	where $\rho_\kappa$ is given in eq. (\ref{Formula:PopularityDist}), and $\mathcal{P}_{\textup{hit}}=\sum_{\kappa=1}^{N}\rho_\kappa$, which is the hitting probability.
\end{theorem}
\begin{IEEEproof}
	See Appendix \ref{appendix:ModesProbe}. 
\end{IEEEproof}
Using the above result, we can obtain the probability that an arbitrary node operates in the transmitting mode, which is given by the following Corollary. This probability will be used in the next sections. 
\begin{cor}
	\label{Corollary: HD transmitting probability corollary}
	The probability that an arbitrary node operates in the transmitting mode, denoted by $\mathcal{P}_{\textup{TX}}$, is 	
	\begin{align}
	\label{Formula: Probe Transmitting}
	\mathcal{P}_{\textup{TX}} = \frac{1}{N} \sum_{\kappa=1}^{N}\left(1-\left(1-\rho_\kappa\right)^{N-1}\right).
	\end{align}
\end{cor}
\begin{IEEEproof}
	A transmitting node should operate either in SR-HDTX, HDTX, or FDTR mode, which means $	\mathcal{P}_\textup{TX} = \mathcal{P}_\textup{SR-HDTX} + \mathcal{P}_\textup{HDTX} + \mathcal{P}_{\rm FDTR}.$ By substituting eqs. (\ref{Formula: Probe SR-HDTX}) and (\ref{Formula: Probe HDTX}) in $\mathcal{P}_{\rm TX}$, we can get the final expression given in eq. (\ref{Formula: Probe Transmitting}).
	%	\begin{equation}
	%	\label{Formula: SR-HDTX+HDTX}
	%	\mathcal{P}_\textup{TX} = \mathcal{P}_\textup{SR-HDTX} + \mathcal{P}_\textup{HDTX} + \mathcal{P}_{\rm FDTR}.
	%	\end{equation}
	
\end{IEEEproof}

%%%%%%%%%%%%%%%%%%%%%%%%%%%%%%%%
% Probabilities Evaluations
%%%%%%%%%%%%%%%%%%%%%%%%%%%%%%%%
%\begin{figure}[t]
%	\centering
%	\includegraphics[width=0.5 \textwidth]{AllProbes.eps}
%	\caption{Probability of all cases versus $N$ for any arbitrary node, with $\gamma_r=1.2$, $m=1000$.}
%	\label{Fig: All Cases Probes}
%\end{figure}

%%%%%%%%%%%%%%%%%%%%%%%%%%%%%%%%%
\section{Success Probability Analysis} \label{sec:Success Analysis}
%%%%%%%%%%%%%%%%%%%%%%%%%%%%%%%%%
An important intermediate step for the interference modeling and success probability analysis is the derivation of the distance distributions associated with the serving distance $Z$ and the interfering distance $W$ shown in Fig. \ref{Fig:RelevantDistances}. 

\subsection{Distance Distributions}
\label{subsec: Distance distribution}
First, we aim to derive the PDF of distance $Z$ between the serving node and the receiver of interest. It is worth noting that probabilities of the operating modes, which are obtained in Theorem \ref{Theorem:OptimalCaching}, determine the number of transmitters as mentioned in Corollary \ref{Corollary: HD transmitting probability corollary}. Since these probabilities do not depend upon the distances between D2D pairs and the radius of the disk $\mathcal{R}$, the serving node and the receiver of interest are both chosen uniformly at random from amongst $N$ nodes. Let $N_t$ be the number of concurrently transmitting nodes, $Z_0=\left\|\textbf{y}_0-\textbf{x}_0 \right\|$ be the distance of the serving link, $\mathcal{W} = \{ W_i | W_i = \left\|  \textbf{y}_i - \textbf{x}_0 \right\| \}_{i=1:(N_t-1)}$ be the set of distances from interfering nodes to the receiver of interest, and $\mathcal{Z} = \{ Z_i | Z_i = \left\|  \textbf{y}_i - \textbf{x}_i \right\| \}_{i=1:(N_t-1)}$ be the set of distances from the interfering nodes to their respective receivers. We can infer that the distance of the receiver of interest $\left\| \textbf{x}_0\right\|$ is a common factor in $Z_0$ and $\mathcal{W}$, and distance of the interfering node $\left\|\textbf{y}_i\right\|$ is a common factor in $\mathcal{W}$ and $\mathcal{Z}$. By conditioning on $\left\|\textbf{x}_0\right\|$, the elements in $\mathcal{W}$ and $Z_0$ become independent, and by conditioning on $\left\|\textbf{y}_i\right\|$, the elements in $\mathcal{W}$ and $\mathcal{Z}$ become independent. This observation will facilitate the analysis of the Laplace transform of the interference field, which is a key to the analysis of the success probability. The following Lemma provides the conditional PDF of distances $Z_0$ and $Z_i$.
\begin{lemma}
	\label{lemma: Dist. Dist of z}
	The conditional PDF of the distance $Z_\lambda$ for a given $q \in \{\left\|\textbf{x}_0\right\|, \left\|\textbf{y}_i\right\|\}$ can be written as 
	\begin{equation}
	\label{Formula: PDF of Z}
	{f_{Z_\lambda}}(z_\lambda|q) = \left\{ \begin{array}{l}
	\begin{array}{*{20}{c}}
	{{f_{Z_\lambda,1}}({z_\lambda}|{q})}&{;0 \le z_\lambda \le {\mathcal{R}} - {q}}
	\end{array}\\
	\begin{array}{*{20}{c}}
	{{f_{Z_\lambda,2}}({z}_\lambda|{q})}&{;{\mathcal{R}} - {q} < z_\lambda \le {\mathcal{R}} + {q}}
	\end{array}
	\end{array} \right.,
	\end{equation}
	where, ${q} = \left\{ \begin{array}{l}
	\begin{array}{*{20}{c}}
	{\left\|\textbf{x}_0\right\|}&{;\lambda=0}
	\end{array}\\
	\begin{array}{*{20}{c}}
	{\left\|\textbf{y}_i\right\|}&{;\lambda=i}
	\end{array}
	\end{array} \right.$, $f_{Z_\lambda,1}(z_\lambda|q) = \frac{2z_\lambda}{\mathcal{R}^2}$, $f_{Z_\lambda,2}(z_\lambda|q)=\frac{2z_\lambda}{\pi {\mathcal{R}}^2}\arccos \left(\frac{z_\lambda^2+q^2-\mathcal{R}^2}{2q z_\lambda}\right)$.
\end{lemma}
\begin{IEEEproof}
	The proof is available in ~\cite[Theorem 2.3.6]{DistanceDistributionsBook}.
\end{IEEEproof}
Now, we need to derive the PDF of distance $W_i$. Similar to the previous Lemma, the elements $\left\|\textbf{x}_0\right\|$ and $\left\|\textbf{y}_i\right\|$ are common factors for $\mathcal{W}$ and $\mathcal{Z}$. Conditioning on $t=\left\|\textbf{y}_i\right\|$ and  $v=\left\|\textbf{x}_0\right\|$ is sufficient to get a set of i.i.d. distances for the elements in $\mathcal{W}$ and $\mathcal{Z}$. The following Lemma provides the conditional PDF of the distance $W_i$.
\begin{lemma}
	\label{theorem: distance dist. of di}
	The conditional PDF of the distance $W_i$ for given $v$ and $t$, denoted by $f_{W_i} \left(w_i |t,v\right)$, is given by 
	\begin{equation}
	\label{Formula: Interf Dista. Dist}
	f_{W_i}\left(w_i|v,t\right) = \frac{1}{\pi} \frac{w_i/(vt)}{\sqrt{1-\left(\frac{v^2+t^2-w_i^2}{2vt}\right)^2}}, \quad \left| v-t \right| <w_i < v+t.
	\end{equation}
\end{lemma}

\begin{IEEEproof}
	The proof is available in ~\cite{MehrnazBPP}. 	
\end{IEEEproof}
For the PDFs of distances $V$ and $T$, it can be easily shown that: $f_{V}(v) = \frac{2v}{\mathcal{R}^2}$ and $f_{T}(t) = \frac{2t}{\mathcal{R}^2}$.

%%%%%%%%%%%%%%%%%%%%%%%%%%%%%%%%%%% 
\subsection{Laplace Transform of the Interference Distribution}
\label{subsec: Laplace transform}
%%%%%%%%%%%%%%%%%%%%%%%%%%%%%%%%%%%
By using the distance distributions obtained in subsection \ref{subsec: Distance distribution}, we aim to obtain the Laplace transform of the interference field through the following Lemma.

\begin{lemma}
\label{Lemma:Laplace Transform}
%The Laplace transform of the interference field at the receiver of interest, by considering the operating mode $\delta \in \{\textup{HDRX, FDTR}\}$ denoted by $\mathcal{L}_{\mathcal{I}, \delta}\left(s\right)$, is given by
The Laplace transform of the interference distribution at the receiver of interest denoted by $\mathcal{L}_{\mathcal{I}, \delta}\left(s\right)$, is
\begin{align}
\mathcal{L}_{\mathcal{I}, \delta}\left(s\right) =  \int_{0}^{\mathcal{R}}\int_{0}^{\mathcal{R}} \mathcal{M}_{\delta} (v,t) \frac{4 v t}{\mathcal{R}^4}\textup{d}v \textup{d}t,
\end{align}
where $\mathcal{M}_{\delta}(v,t)$ is given by eq.  (\ref{Formula: M(v,t) HD}), ${\mathcal{F}_{\delta}(v,t,z_0)}={ \int_{\left|v-t\right|  }^{\left|v+t\right|} \int_{0}^{\mathcal{R}-t} {\mathcal{J}_{\delta}} f_{Z_i,1}(z_i|t)f_{W_i}(w_i|v,t) \textup{d}z_i\textup{d}w_i}$, ${\mathcal{G}_{\delta}(v,t,z_0)} = {\int_{\left|v-t\right|  }^{\left|v+t\right|} \int_{\mathcal{R}-t}^{\mathcal{R}+t} \mathcal{J}_{\delta}f_{Z_i,2}(z_i|t)f_{W_i}(w_i|v,t) \textup{d}z_i \textup{d}w_i}$, and $\mathcal{J}_{\delta} = \frac{\exp (-s \mathbbm{1}_\delta \beta z_0^{\alpha}  )}{1+s z_i^{\alpha} w_i^{-\alpha}}$.
\begin{figure*}[t!]
The $\mathcal{M}_{\delta} (v,t) $ is given by
\begin{align}\label{Formula: M(v,t) HD}
 \begin{cases}
 \left( \mathcal{F}_\delta(v,t,z_0) +\mathcal{G}_\delta(v,t,z_0)\vphantom{\int_{\left|v-t\right|  }^{\left|v+t\right|}} \right)^{N_t-1} 
&\delta=\textup{HDRX}\\
 \int_{0}^{\mathcal{R}-v} \left( \mathcal{F}_{\delta}(v,t,z_0) +\mathcal{G}_{\delta}(v,t,z_0)\vphantom{\int_{\left|v-t\right|  }^{\left|v+t\right|}} \right)^{N_t-1} f_{Z_0,1}(z_0|v) {\rm d}z_0 + \int_{\mathcal{R}-v}^{\mathcal{R}+v} \left( \mathcal{F}_{\delta}(v,t,z_0) +\mathcal{G}_{\delta}(v,t,z_0)\vphantom{\int_{\left|v-t\right|  }^{\left|v+t\right|}} \right)^{N_t-1} f_{Z_0,2}(z_0|v){\rm d}z_0&\delta=\textup{FDTR}\\
\end{cases}
  \end{align}

%\begin{align}\label{Formula: M(v,t) HDRX}
% &\mathcal{M} (v,t) =  \left( \mathcal{F}_\delta(v,t,z_0) +\mathcal{G}_\delta(v,t,z_0)\vphantom{\int_{\left|v-t\right|  }^{\left|v+t\right|}} \right)^{N_t-1} 
%  \end{align}
%
%
%
% \begin{align}\label{Formula: M(v,t) HD}
%&\mathcal{M} (v,t) =  \int_{0}^{\mathcal{R}-v} \left( \mathcal{F}(v,t,z_0) +\mathcal{G}(v,t,z_0)\vphantom{\int_{\left|v-t\right|  }^{\left|v+t\right|}} \right)^{N_t-1} f_{Z_0,1}(z_0|v) {\rm d}z_0 + \int_{\mathcal{R}-v}^{\mathcal{R}+v} \left( \mathcal{F}(v,t,z_0) +\mathcal{G}(v,t,z_0)\vphantom{\int_{\left|v-t\right|  }^{\left|v+t\right|}} \right)^{N_t-1} f_{Z_0,2}(z_0|v){\rm d}z_0,
%%\\& \textup{where,} \notag\\&
%%{\mathcal{F}(v,t,z_0)}={ \int_{\left|v-t\right|  }^{\left|v+t\right|} \int_{0}^{\mathcal{R}-t} {\mathcal{J}} f_{Z_i,1}(z_i|t)f_{W_i}(w_i|v,t) \textup{d}z_i\textup{d}w_i}, \quad {\mathcal{G}(v,t,z_0)} = {\int_{\left|v-t\right|  }^{\left|v+t\right|} \int_{\mathcal{R}-t}^{\mathcal{R}+t} \mathcal{J}f_{Z_i,2}(z_i|t)f_{W_i}(w_i|v,t) \textup{d}z_i \textup{d}w_i}. \notag
% \end{align}
%
%$\delta \in \{\textup{HDRX, FDTR}\}$
%  \begin{align}\label{Formula: M(v,t) HDRX}
% &\mathcal{M} (v,t) =  \left( \mathcal{F}(v,t,z_0) +\mathcal{G}(v,t,z_0)\vphantom{\int_{\left|v-t\right|  }^{\left|v+t\right|}} \right)^{N_t-1} 
%  \end{align}
\noindent\rule{18.5cm}{0.4pt}
\end{figure*}
\end{lemma}
\begin{IEEEproof}
	See Appendix \ref{appendix:Laplace Transform}.
\end{IEEEproof}
We use Laplace transform of the interference distribution to obtain the success probability, which is formally defined next.
%\subsection{Success Probability}
%\label{subsec:coverage probability}
\begin{ndef}
[Success Probability]	
\label{Defin:CoverageProbe}
It is the probability that an arbitrary node has its desired content in its own cache or can receive it successfully from a nearby device over a D2D link. In the latter case, the transmission is said to be successful only if the received $\sir$ exceeds some predefined threshold $\theta$.
\end{ndef}
\begin{theorem}
	\label{Theorem:SuccessProbe}
	The success probability for an arbitrary node is denoted by $\textup{P}_{\rm {s}}(N,\gamma_r,\theta)$ and given by
	\begin{align}
	\label{Formula: Final Success Probe}
	\textup{P}_{\rm {s}}(N,\gamma_r,\theta) = \textup{P}_{\rm {s},\rm cache} (N,\gamma_r) + \textup{P}_{\rm {s}, \sir} (N,\gamma_r,\theta),
	\end{align}
 \begin{align*}
\text{where} \quad \textup{P}_{\rm {s},\rm cache}(N,\gamma_r)& = \frac{1}{N} \mathcal{P}_{\rm hit}\\
  \textup{P}_{\rm {s},\sir} (N,\gamma_r,\theta) &= \sum_{n_t = 1}^{N} 	\textup{Q}_{\rm {s}} \left(N,\gamma_r,\theta\right) f_{N_t}(n_t),
\end{align*}
 $\textup{Q}_{\rm {s}} \left(N,\gamma_r,\theta\right) =  \mathcal{P}_{\rm HDRX}\mathcal{L}_{\mathcal{I},\rm HDRX}\left(\theta\right) \notag + \mathcal{P}_{\rm FDTR}\mathcal{L}_{\mathcal{I},\rm FDTR}\left(\theta\right)$, $f_{N_t}(n_t) = \left(\begin{array}{c}N\\ n_t\end{array}\right)(\mathcal{P}_{\rm {TX}})^{n_t}\left(1-\mathcal{P}_{\rm {TX}}\right)^{N-n_t}$, and  $\mathcal{P}_{\rm TX}$ is given by Corollary \ref{Corollary: HD transmitting probability corollary}.
%$$\text{with} \quad \textup{P}_{\rm {s},\rm cache}(N,\gamma_r) = \frac{1}{N} \mathcal{P}_{\rm hit}$$
%$${\text and} \quad \textup{P}_{\rm {s},\sir} (N,\gamma_r,\theta) = \sum_{n_t = 1}^{N} 	\textup{Q}_{\rm {s}} \left(N,\gamma_r,\theta\right) f_{N_t}(n_t)$$		
%	
%	where $\textup{P}_{\rm {s},\rm cache}(N,\gamma_r) = \frac{1}{N} \mathcal{P}_{\rm hit}$,\\ $\textup{P}_{\rm {s},\sir} (N,\gamma_r,\theta) = \sum_{n_t = 1}^{N} 	\textup{Q}_{\rm {s}} \left(N,\gamma_r,\theta\right) f_{N_t}(n_t)$, \\$\textup{Q}_{\rm {s}} \left(N,\gamma_r,\theta\right) =  \mathcal{P}_{\rm HDRX}\mathcal{L}_{\mathcal{I},\rm HDRX}\left(\theta\right) \notag + \mathcal{P}_{\rm FDTR}\mathcal{L}_{\mathcal{I},\rm FDTR}\left(\theta\right)$, $f_{N_t}(n_t) = \left(\begin{array}{c}N\\ n_t\end{array}\right)(\mathcal{P}_{\rm {TX}})^{n_t}\left(1-\mathcal{P}_{\rm {TX}}\right)^{N-n_t}$, and  $\mathcal{P}_{\rm TX}$ is given by Corollary \ref{Corollary: HD transmitting probability corollary}.
\end{theorem}
\begin{IEEEproof}
See Appendix \ref{appendix:SuccessProbe}.
\end{IEEEproof}

%%%%%%%%%%%%%%%%%%%%%%%%%%%
\section{Results and Discussion} \label{sec:NumResults}
%%%%%%%%%%%%%%%%%%%%%%%%%%%
For the popularity distribution, we use Zipf distribution, which is a special case of the Riemann Zeta function and is widely used in the existing literature~\cite{Negin_Mag, cacherequestPPP,HitProbe1, HitProbe2,HitProbe3,malak,MehrnazBPP}. This distribution is defined by $\rho_\kappa = \Upsilon (\kappa,\gamma_r,m) = {\kappa}^{-\gamma_r} \left(\sum_{\eta=1}^{m} {\eta }^{-\gamma_r}\right)^{-1}$. Here, we assume $m=1000$, $\alpha=4$, and $\beta = 10^{-5}$. 
\subsubsection{Impact of $N$}
\begin{figure}[t]
	\centering
	\includegraphics[width=0.44 \textwidth]{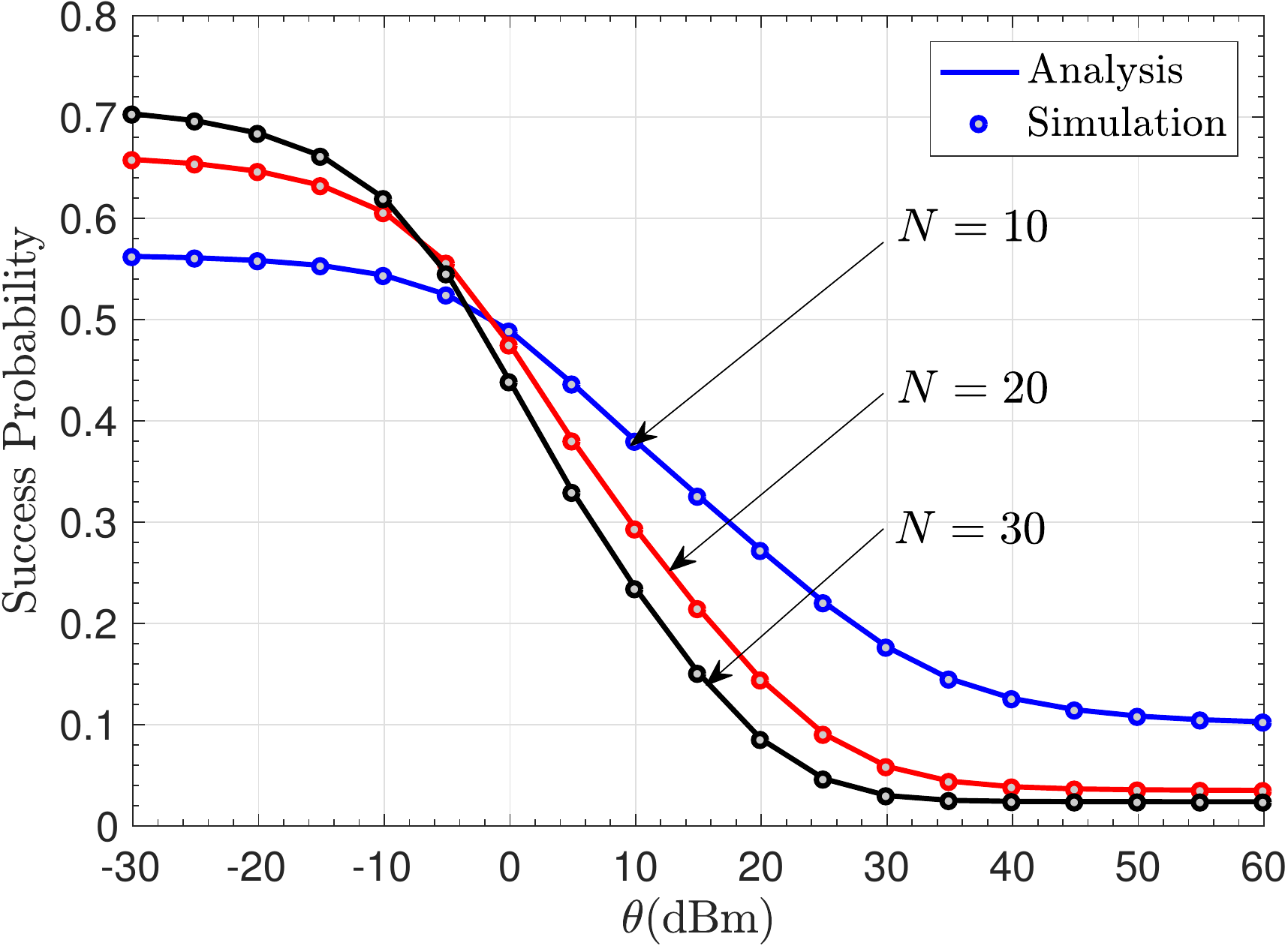}
	\caption{Coverage Probability versus $\sir$ threshold $\theta$ for different values of $N$, and $\gamma_r=1.2$, $\mathcal{R}=30$.}
	\label{Fig:CoverageProbe_N}
\end{figure}

Fig. \ref{Fig:CoverageProbe_N} demonstrates the effects of $N$ on the success probability. Intuitively, increasing $N$ has two conflicting effects: it increases {\em content availability} in the network at the expense of higher interference. Specifically, more users in the network mean higher content availability but also higher interference. Our numerical comparisons reveal that the content availability dominates the performance in the regime of lower target $\sir$ thresholds. This is because in this regime, the complementary cumulative distribution function (CCDF) of the $\sir$ is almost the same for all values of $N$. However, when the threshold increases, the interference starts dominating the performance. The effect is even more pronounced for higher values of $N$. %In other words, one can observe that the curves start getting steeper with increasing value of $N$.

\subsubsection{Impact of Zipf exponent}
\begin{figure}[t]
	\centering
	\includegraphics[width=0.44 \textwidth]{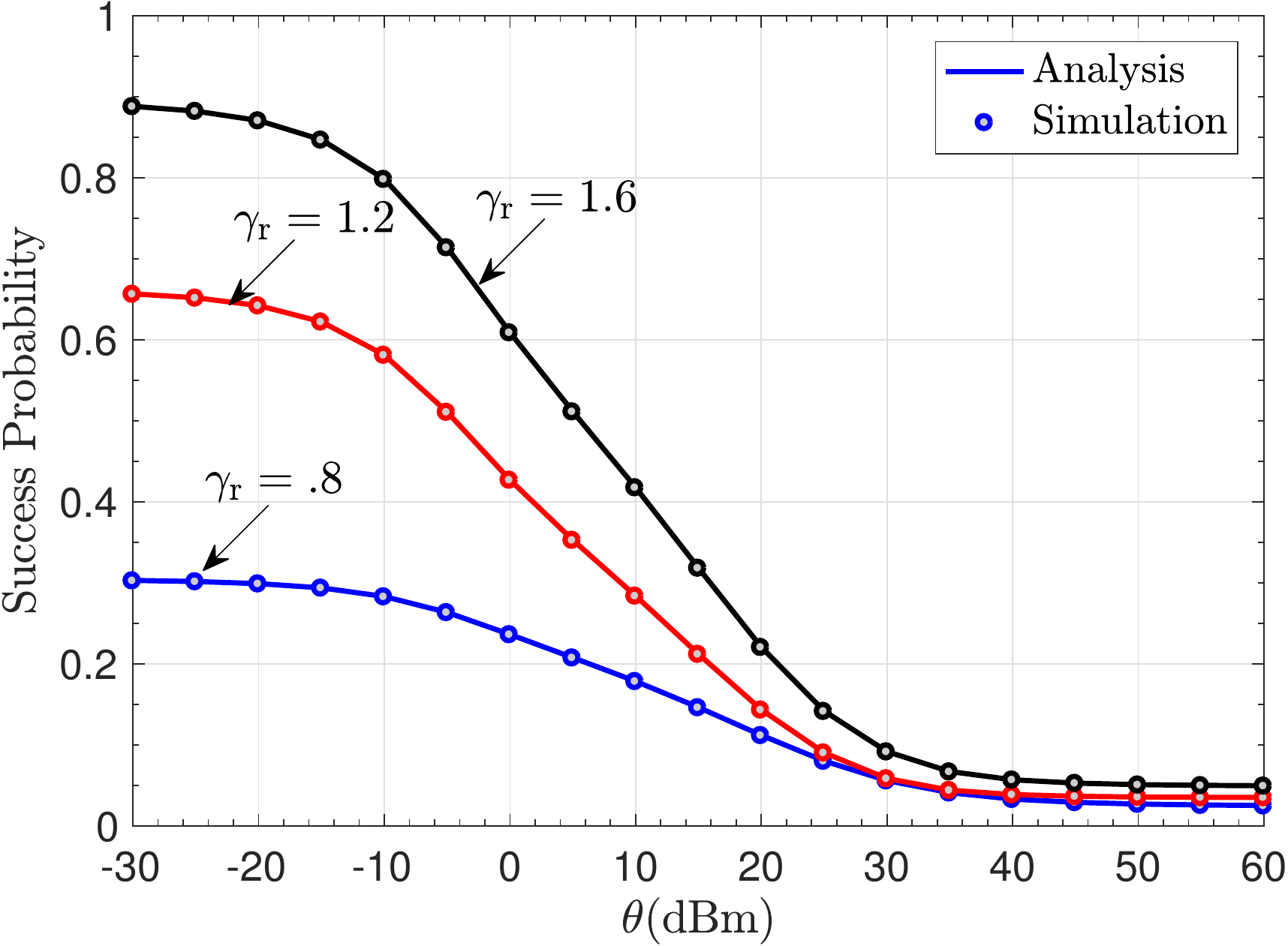}
	\caption{Coverage Probability versus $\sir$ threshold $\theta$ for different values of $\gamma_r$, and $N=20$, $\mathcal{R}=40$.}
	\label{Fig:CoverageProbe_Zipf}
\end{figure}

Fig. \ref{Fig:CoverageProbe_Zipf} illustrates the effect of Zipf exponent $\gamma_r$ on the success probability. Technically, the higher values of the $\gamma_r$ imply more redundancy in the users' demands, that is to say, fewer number of the contents accounts for the majority of requests. Correspondingly, caching performance dominates against $\sir$ distribution, when $\gamma_r$ increases. As was the case above, caching dictates the system performance in lower target $\sir$ thresholds whereas interference dictates the performance at the higher target $\sir$ thresholds.

%%%%%%%%%%%%%%%%%%%%%%%%%%%%%%%%%%%%%%%%%%%%
% Enable this for the operating modes validations
%%%%%%%%%%%%%%%%%%%%%%%%%%%%%%%%%%%%%%%%%%%%
%\begin{figure}[t]
%	\centering
%	\includegraphics[width=0.5 \textwidth]{AllProbes.eps}
%	\caption{Operating Modes Probability for an arbitrary node versus $N$ for $\gamma_r=1.2$.}
%	\label{Fig:OperatingModesProbes}
%\end{figure}

%%%%%%%%%%%%%%%%%%%%%%%%%%%%
\section{Concluding Remarks}
%%%%%%%%%%%%%%%%%%%%%%%%%%%%
In this paper, we have derived closed form expressions for the probabilities of different operating modes that appear in a cache-enabled D2D network in which the users have FD capability. This characterization allowed us to determine the number of D2D devices that transmit concurrently at any given time, which ultimately facilitated network performance analysis in terms of success probability, which additionally depends on the content availability, and $\sir$ distribution. Exploring more realistic caching policies such as random caching policy, will be a useful extension to this work. Moreover, in
this paper and all related works, the channel for the FD links is assumed to be reciprocal and exchanging data on a FD nodes is assumed to be symmetric. FD communications with asymmetric data and dissimilar channel is another worthwhile open problem to pursue.

%%%%%%%%%
\appendix
%%%%%%%%%

\subsection{Proof for Theorem \ref{Theorem:OptimalCaching}} 
\label{appendix:ModesProbe}
From Fig. \ref{Fig: D2DCasesGraph}, we can infer that the probability of occurrence of each case at an arbitrary user $u_\kappa$ depends on two different events: i) request of the user $u_\kappa$ (we denote this event by $\mathcal{A}$ and the probability of this event by $\mathcal{P}_{\Delta}^{a,\kappa
}$) and ii) requests from other users for the content cached in user $u_\kappa$. We denote this event by $\mathcal{B}$ and the probability of this event by $\mathcal{P}_{\Delta}^{b,\kappa}$. The joint probability of both events, i.e., $\textup{P} (\mathcal{A},\mathcal{B})$ gives the probability of the operating mode $\Delta$ for a specific node $u_\kappa$. Since the requests at all users are independent from each other, we can say that $\textup{P} (\mathcal{A},\mathcal{B}) = \textup{P}(\mathcal{A})\textup{P}(\mathcal{B})=\mathcal{P}_{\Delta}^{a,\kappa
}\mathcal{P}_{\Delta}^{b,\kappa
}$. Now, by using the law of total probability, the probability of operating mode $\Delta \in $ \{SR, SR-HDTX, FDTR, BFD, TNFD, HDTX, HDRX, HO\} denoted by $\mathcal{P}_\Delta$ for an arbitrary node can be defined by 
\begin{equation}
\label{P ab}
\mathcal{P}_{\Delta} = \sum_{\kappa=1}^{N} \mathcal{P}_{\Delta}^{a,\kappa} \mathcal{P}_{\Delta}^{b,\kappa} \mathcal{P}_{u_\kappa},
\end{equation}
where, $\mathcal{P}_{u_\kappa} = \frac{1}{N}$ is the probability of choosing an arbitrary user among $N$ users uniformly at random. Due to lack of space, we provide the proof for the FDTR mode, however, the approach remains the same for the other modes as well. Now, let us define two binary random variables $\mathcal{X}_\kappa$ and $\mathcal{H}_{\mu,\kappa}$ for $u_\kappa$ as follows.
\begin{equation}
{\mathcal{X}_\kappa} = \left\{ \begin{array}{l}
\begin{array}{*{20}{c}}
0&{;{ u_\kappa \textup{ cannot find its desired content}}}
\end{array}\\
\begin{array}{*{20}{c}}
1&{;{ u_\kappa \textup{ can find its desired content,}}}
\end{array}
\end{array} \right.
\end{equation}
\begin{equation}
{\mathcal{H}_{\mu,\kappa}} = \left\{ \begin{array}{l}
\begin{array}{*{20}{c}}
0&{;{ u_\mu \textup{ does not demands for content $c_\kappa$}}}
\end{array}\\
\begin{array}{*{20}{c}}
1&{;{ u_\mu \textup{ demands for content $c_\kappa$.}}}
\end{array}
\end{array} \right.
\end{equation}
The probability $\Pr \left(\mathcal{X}_\kappa=1\right)$ is equivalent to the situation that $u_\kappa$ demands for a content, which is cached by other node in its vicinity, i.e., $\Pr \left(\mathcal{X}_\kappa=1\right) = \mathcal{P}_{\rm hit}-\rho_\kappa$, which corresponds to the parameter $\mathcal{P}_{\rm FDTR}^{a,\kappa}$, i.e.,
\begin{equation}
\label{P_a}
\mathcal{P}_{\rm FDTR}^{a,\kappa} = \Pr \left(\mathcal{X}_\kappa=1\right),
\end{equation}
and the probability $\Pr \left(\mathcal{H}_{\mu, \kappa}=0\right)$ is equivalent to
\begin{equation}
\label{Pr (H=0)}
\Pr \left(\mathcal{H}_{\mu,\kappa}=0\right) = 1-\rho_\kappa.
\end{equation}
Now, the parameter $\mathcal{P}_{\rm FDTR}^{b,\kappa}$ is equivalent to the probability that there is at least one node that demands content $c_\kappa$, hence 
\begin{align}
\label{P_b}
\mathcal{P}_{\rm FDTR}^{b,\kappa} = &1- \Pr \bigg(  \bigcup_{\mu = 1,\mu \ne \kappa}^{N} \mathcal{H}_{\mu,\kappa}=0  \bigg) \notag \\ \mathop  = \limits^{(a)} & 1- \prod_{\mu=1, \mu \ne \kappa}^{N} \Pr \left( \mathcal{H}_{\mu,\kappa} = 0\right) \notag \\ \mathop  = \limits^{(b)} & 1- \left(1-\rho_\kappa\right)^{N-1},
\end{align}
where (a) follows the fact that the requests at all users are independent from each other and (b) follows directly using the eq. (\ref{Pr (H=0)}). By substituting the eqs. (\ref{P_a}) and (\ref{P_b}) in eq. (\ref{P ab}), we can get the final expression in eq. (\ref{Formula: Probe FDTR}).

%%%%%%%%%%%%%%%%%%%%%%%%%%%%%%%%%
\subsection{Proof for Lemma \ref{Lemma:Laplace Transform}}
\label{appendix:Laplace Transform}
%%%%%%%%%%%%%%%%%%%%%%%%%%%%%%%%%%
The  Laplace transform of the interference $\mathcal{I} = \sum_{\mathbf{y}_i \in \Phi \backslash \mathbf{y}_0}  \left( h_i Z_i^{\alpha}W_i^{-\alpha} + \mathbbm{1}_\delta \beta Z_0^{\alpha} \right)$ is $\mathcal{L}_{\mathcal{I}, \delta}(s)$ 
%
%According to Fig. \ref{Fig:RelevantDistances}, denoting $Z_0 = \left\|  {\rm \mathbf{y}}_0 -{\rm \mathbf{x}}_0  \right\|$, $Z_i = \left\|  {\rm \mathbf{y}}_i -{\rm \mathbf{x}}_i  \right\|$, and $W_i = \left\|  {\rm \mathbf{y}}_i -{\rm \mathbf{x}}_0  \right\| $, the Laplace transform of the interference $\mathcal{I} = \sum_{\mathbf{y}_i \in \Phi \backslash \mathbf{y}_0}  \left( h_i Z_i^{\alpha}W_i^{-\alpha} + \mathbbm{1}_\delta \beta Z_0^{\alpha} \right)$ is
\begin{align*}
\begin{split}
%& \mathcal{L}_{\mathcal{I}, \delta}(s) =   \mathbb{E}\left[e^{-s\mathcal{I}}\right]  \\=& 
=&    \mathbb{E} \bigg[ \exp \bigg( -s\sum_{\mathbf{y}_i \in \Phi \backslash \mathbf{y}_0}  \left( h_i Z_i^{\alpha}W_i^{-\alpha} + \mathbbm{1}_\delta \beta Z_0^{\alpha} \right) \bigg) \bigg] \\=& \mathbb{E} \bigg[ \prod_{\mathbf{y}_i \in \Phi \backslash \mathbf{y}_0} \exp \left( -s  \bigg( h_i Z_i^{\alpha}W_i^{-\alpha} + \mathbbm{1}_\delta \beta Z_0^{\alpha} \bigg)   \right) \bigg] \\\mathop = \limits^{(a)} & \mathbb{E} \bigg[ \prod_{\mathbf{y}_i \in \Phi \backslash \mathbf{y}_0} \frac{1}{1+s Z_i^{\alpha}W_i^{-\alpha}} e^{-s \mathbbm{1}_\delta \beta Z_0^{\alpha}} \bigg] \notag \\ \mathop = \limits^{(b)} & \mathbb{E}  \bigg[ \bigg( \underbrace { \int_{\left|v-t\right|  }^{\left|v+t\right|} \int_{0}^{\mathcal{R}-t} {\mathcal{J}_{\delta}} f_{Z_i,1}(z_i|t)f_{W_i}(w_i|v,t) \textup{d}z_i\textup{d}w_i}_{\mathcal{F}_{\delta}(v,t,z_0)}  \\ &
+  \underbrace {\int_{\left|v-t\right|  }^{\left|v+t\right|} \int_{\mathcal{R}-t}^{\mathcal{R}+t} \mathcal{J}_{\delta}f_{Z_i,2}(z_i|t)f_{W_i}(w_i|v,t) \textup{d}z_i \textup{d}w_i}_{\mathcal{G}_{\delta}(v,t,z_0)} \bigg)^{N_t-1} \bigg],
% \\ \mathop = \limits^{(c)} & \mathbb{E} \Bigg[ \int_{0}^{\mathcal{R}-v} \left( \mathcal{F}(v,t,z_0) +\mathcal{G}(v,t,z_0)\vphantom{\int_{\left|v-t\right|  }^{\left|v+t\right|}} \right)^{N_t-1} f_{Z_0,1}(z_0|v) {\rm d}z_0  \\+&  \int_{\mathcal{R}-v}^{\mathcal{R}+v} \left( \mathcal{F}(v,t,z_0) +\mathcal{G}(v,t,z_0)\vphantom{\int_{\left|v-t\right|  }^{\left|v+t\right|}} \right)^{N_t-1} f_{Z_0,2}(z_0|v){\rm d}z_0 \Bigg],
\end{split}
\end{align*}
where $\mathcal{J}_{\delta}= \begin{cases} \frac{1}{1+s z_i^{\alpha} w_i^{-\alpha}}&\delta=\textup{HDRX}\\
 \frac{\exp (-s  \beta z_0^{\alpha}  )}{1+s z_i^{\alpha} w_i^{-\alpha}}&\delta=\textup{FDTR}\end{cases}$.  Here $Z_0 = \left\|  {\rm \mathbf{y}}_0 -{\rm \mathbf{x}}_0  \right\|$, $Z_i = \left\|  {\rm \mathbf{y}}_i -{\rm \mathbf{x}}_i  \right\|$, and $W_i = \left\|  {\rm \mathbf{y}}_i -{\rm \mathbf{x}}_0  \right\| $. Step (a) follows from expectation with respect to $h_i \sim \exp(1)$ and step (b) follows from expectation with respect to $Z_i$ and $W_i$ using  the PDFs of $Z_i$ and $W_i$ given by eqs. (\ref{Formula: PDF of Z}) and (\ref{Formula: Interf Dista. Dist}). From this step, the final expression is obtained by taking expectation with respect to $Z_0$, $V$, and $T$.

\subsection{Proof for Theorem \ref{Theorem:SuccessProbe}}
\label{appendix:SuccessProbe}
%%%%%%%%%%%%%%%%%%%%%%%%%%%%%%%%%%
The success probability contains two different parts. The first part relates to the caching aspects, and the second part relates to the $\sir$ distribution. Both parts depend on the number of users $N$, skew exponent $\gamma_r$, and the $\sir$ threshold $\theta$. We denote the first and second parts as $\textup{P}_{\rm {s}}(N,\gamma_r,\theta)$ and $\textup{P}_{\rm {s}, \sir} (N,\gamma_r,\theta)$, respectively. Hence, the success probability denoted by $\textup{P}_{\rm s} (N,\gamma_r,\theta)$ can be defined as in eq. (\ref{Formula: Final Success Probe}). From the observations in Fig. \ref{Fig: D2DCasesGraph}, we infer that an arbitrary node in the cases SR and SR-HDTX, can capture its desired content directly through its own cache. Hence the proof for caching part in success probability, namely $\textup{P}_{\rm {s},\rm cache}(N,\gamma_r)$ is straightforward:
\begin{align}
\label{Formula:FirstPart Success}
\textup{P}_{\rm {s},\rm cache}(N,\gamma_r) = & \mathcal{P}_{\rm SR} + \mathcal{P}_{\rm SR-HDTX} \notag\\ \mathop=^{(a)} & \frac{1}{N} \sum_{\kappa=1}^{N} \rho_\kappa=\frac{1}{N} \mathcal{P}_{\rm hit},
\end{align}
where (a) follows substituting $\mathcal{P}_{\rm SR}$ and $\mathcal{P}_{\rm SR-HDTX}$, which are given respectively in eqs. (\ref{Formula: Probe SR}) and (\ref{Formula: Probe SR-HDTX}). The eq. (\ref{Formula:FirstPart Success}) completes the proof for the first part. We denote CCDF of the $\sir_\delta$ by $\textup{P}(\sir_\delta \ge \theta)$ and it can be easily shown that
\begin{equation}
\label{Formula:CCDF of SIR}
\textup{P}(\sir_\delta \ge \theta) = \mathcal{L}_{\mathcal{I},\delta}\left(\theta\right).
\end{equation}
The success probability for the receiver of interest denoted by $\textup{Q}_{\rm {s}} \left(N,\gamma_r,\theta\right)$ depends on its operation mode, which is either HDRX or FDTR. Hence, we have 
\begin{align}
\label{Formula:Qs}
\textup{Q}_{\rm {s}} \left(N,\gamma_r,\theta\right) = & \sum_{\delta \in \{\rm HDRX, FDTR \}  } \mathcal{P}_{\delta}  \textup{P}\left(\sir_{\delta} \ge \theta\right) \notag \\ \mathop =^{(a)}& \mathcal{P}_{\rm HDRX}\mathcal{L}_{\mathcal{I},\rm HDRX}\left(\theta\right) + \mathcal{P}_{\rm FDTR}\mathcal{L}_{\mathcal{I},\rm FDTR}\left(\theta\right),
\end{align}
where (a) follows substitution from eq. (\ref{Formula:CCDF of SIR}). The expression in eq. (\ref{Formula:Qs}) contains the parameter $N_t$, which is the number of transmitters and is random. From the Fig. \ref{Fig: D2DCasesGraph}, we can infer that the distribution of the random variable $N_t$ depends on the probability of the transmitting operation modes, which is defined in Corollary \ref{Corollary: HD transmitting probability corollary}, i.e., $\mathcal{P}_{\rm TX}$. The number of transmitters $N_t$ is a Binomial random variable and its PMF is $f_{N_t}(n_t) = \bigg(\begin{array}{c}N\\ n_t\end{array}\bigg)(\mathcal{P}_{\rm {TX}})^{n_t}\left(1-\mathcal{P}_{\rm {TX}}\right)^{N-n_t}$. Now, the final expression for the second part of the success probability $\textup{P}_{\rm {s}, \sir} (N,\gamma_r,\theta)$ can be obtained by taking expectation of $\textup{Q}_{\rm {s}} \left(N,\gamma_r,\theta\right)$ over the random variable $N_t$, i.e., 
\begin{align}
\label{Formula:SecondPart Success}
\textup{P}_{\rm {s},\sir} \left(N,\gamma_r,\theta \right) = &  \mathbb{E}_{N_t} \left[ \textup{Q}_{\rm {s}} \left(N,\gamma_r,\theta\right)  \right] \notag \\=& \sum_{n_t=1}^{N} \textup{Q}_{\rm {s}} \left(N,\gamma_r,\theta\right) f_{N_t}(n_t),
\end{align}
which completes the proof for the second part. Finally, substituting eqs. (\ref{Formula:FirstPart Success}) and (\ref{Formula:SecondPart Success}) in eq. (\ref{Formula: Final Success Probe}) completes the proof.

%\IEEEQED

\bibliographystyle{IEEEtran}
%\bibliography{Hokies}
\bibliography{ICC2018FinalVersion}

\end{document}